\documentclass[aps,prl,twocolumn,superscriptaddress,showpacs,floatfix,preprintnumbers]{revtex4-1}

\usepackage{epsfig}
\usepackage{wrapfig}
\usepackage{graphicx}
\usepackage{bm}
\usepackage{amssymb}
\usepackage{amsmath}
\usepackage{amsfonts}
\usepackage[dvipsnames]{xcolor}
\usepackage{subfigure}
\usepackage{dcolumn}
\usepackage[makeroom]{cancel}
\usepackage{float}
\usepackage{xcolor}

\usepackage[utf8]{inputenc}

\usepackage{amsmath}

\usepackage{amssymb}
\usepackage{amsthm}
\usepackage{amscd, amsfonts, mathrsfs}
\usepackage{cases}
\usepackage{verbatim} 
\usepackage{epsf}
\usepackage[bookmarksnumbered,pdfpagelabels=true,plainpages=false,colorlinks=true,
            linkcolor=black,citecolor=black,urlcolor=Blue]{hyperref}

\usepackage{orcidlink}
\theoremstyle{plain}

\theoremstyle{definition}

\allowdisplaybreaks

\usepackage{appendix}

\newcommand{\al}{\alpha}
\newcommand{\ga}{\gamma}
\newcommand{\Ga}{\Gamma}
\newcommand{\de}{\delta}
\newcommand{\ep}{\varepsilon}
\newcommand{\la}{\lambda}

\newcommand{\si}{\sigma}

\newcommand{\om}{\omega}

\newcommand{\pa}{\partial}

\newcommand{\be}{\begin{equation}}
\newcommand{\ee}{\end{equation}}
\newcommand{\bea}{\begin{aligned}}
\newcommand{\eea}{\end{aligned}}

\newcommand{\bml}{\begin{subequations}}
\newcommand{\eml}{\end{subequations}}

\newcommand{\bbm}{\begin{bmatrix}}
\newcommand{\ebm}{\end{bmatrix}}
\newcommand{\bvm}{\begin{vmatrix}}
\newcommand{\evm}{\end{vmatrix}}

\begin{document}

\title{Symmetry energy dependence of the bulk viscosity of nuclear matter}

\date{\today}

\author{Yumu Yang\,\orcidlink{0009-0001-8979-9343}}
\email{yumuy2@illinois.edu}
\affiliation{Illinois Center for Advanced Studies of the Universe \& Department of Physics, 
University of Illinois Urbana-Champaign, Urbana, IL 61801, USA}

\author{Mauricio Hippert\,\orcidlink{0000-0001-5802-3908}}
\email{hippert@cbpf.br}
\affiliation{Centro Brasileiro de Pesquisas Físicas, 
Rio de Janeiro, RJ, 22290-180, Brasil}
\affiliation{Universidade do Estado do Rio de Janeiro, Rio de Janeiro, RJ, 20550-013, Brasil}

\author{Enrico Speranza\,\orcidlink{0000-0003-3076-6958}}
\email{enrico.speranza@cern.ch}
\affiliation{Theoretical Physics Department, CERN, 1211 Geneva 23, Switzerland}

\author{Jorge Noronha\,\orcidlink{0000-0002-9817-0272}}
\email{jn0508@illinois.edu}
\affiliation{Illinois Center for Advanced Studies of the Universe \& Department of Physics, 
University of Illinois Urbana-Champaign, Urbana, IL 61801, USA}

\preprint{CERN-TH-2025-078}
\begin{abstract}
We clarify how the weak-interaction-driven bulk viscosity $\zeta$ and the bulk relaxation time $\tau_\Pi$ of neutrino-transparent $npe$ matter depend on the nuclear symmetry energy. We show that, at saturation density, the equation-of-state dependence of these transport quantities is fully determined by the experimentally constrained nuclear symmetry energy $S$ and its slope $L$. Variations of $L$ can change the bulk viscosity by orders of magnitude, which can affect \emph{both} the dissipative and the conservative tidal response 
of neutron stars.
This suggests that both conservative and dissipative effects encoded in the gravitational-wave signatures of binary neutron star inspirals may help constrain nuclear symmetry energy properties.   
\end{abstract}

\maketitle

{\it Introduction} --- 
Due to electric charge neutrality, neutron star matter exhibits isospin asymmetries much larger than those typically found in nuclei, with proton fractions around $\lesssim 10\%$ \cite{norman1997compact}. Thus, the symmetry energy $E_{\textrm{sym}}$, which in nuclear physics quantifies the energy cost of an isospin asymmetry, is crucial for the structure of neutron stars \cite{Lattimer:2000nx}. As a function of the baryon density, $n_B$, the symmetry energy is often parameterized by its value $S$ and its slope $L$ at nuclear saturation density, $n_\mathrm{sat}$, both of which play an important role in the neutron star equation of state \cite{MUSES:2023hyz}. For an in-depth, comprehensive review on nuclear symmetry energy, see Ref.\ \cite{Lattimer:2023rpe}.

Constraints on the nuclear symmetry energy have been derived from both experimental and theoretical efforts. Experimental constraints are obtained, for example, through nuclear mass fitting \cite{Dutra:2012mb, Dutra:2014qga, Tagami:2022xvs, Kortelainen:2010hv} and neutron skin thickness measurements \cite{Steiner:2004fi, Reed:2021nqk, Brown:2000pd, Typel:2001lcw, Xu:2020fdc, Zhang:2013wna}, while theoretical constraints can be obtained through chiral effective field theory (EFT) \cite{Drischler:2020hwi, Tews:2024owl}. While $S$ remains relatively well-constrained, there is ongoing discussion on the value of $L$. Of particular interest are the recent parity-violating electron scattering neutron skin
experiments of ${}^{208}$Pb (PREX-I and PREX-II) \cite{PREX:2021umo} and ${}^{48}$Ca (CREX) \cite{Lattimer:2023rpe}. While most estimates lie around $L \approx 50$ MeV 
\cite{Li:2019xxz,Fan:2014rha,Reinhard:2021utv}, 
PREX-I and II and CREX measurements 
suggest $L=121 \pm 47$ MeV and $L = -5 \pm 40$ MeV, respectively, to 68$\%$ confidence \cite{Lattimer:2023rpe}. There have been efforts in accounting for the results from PREX \cite{Furnstahl:2001un, Danielewicz:2013upa} and reconciling both PREX and CREX results, see \cite{Zhang:2022bni, Reinhard:2022inh, Lattimer:2023rpe}. In particular, Ref.\ \cite{Lattimer:2023rpe} shows that nuclear interactions optimally satisfying both measurements imply $L=53\pm 13$ MeV. Studies on the importance of $L$ in neutron stars, showing for instance that it has a large impact on the stellar radius and tidal deformability, can be found in Refs.\ \cite{Lattimer:2000nx, Yagi:2013awa, Yagi:2015pkc, De:2018uhw, Raithel:2018ncd}. Furthermore, in the post-merger phase of binary neutron star collisions, Ref.\ \cite{Most:2021ktk} found that the amount of dynamically ejected mass increases with $L$, while gravitational wave emission is mostly insensitive to variations of this quantity.

In a previous work \cite{Yang:2023ogo}, we proposed that independent information on the symmetry energy may be obtained by investigating bulk-viscous transport coefficients  of neutron star matter in binary neutron star coalescence.
There, bulk viscosity is expected to emerge as a consequence of weak-interaction-driven flavor equilibration dynamics \cite{Sawyer:1989dp}. Ref.\ \cite{Yang:2023ogo} showed that, at least for equations of state (EOS) derived using relativistic mean-field (RMF) models, the bulk viscosity transport coefficient $\zeta$ of matter composed of protons $p$,
neutrons $n$, and electrons $e$ ($npe$ matter) can be sensitive to changes in $S$ and $L$. 

In this paper, we work out in detail how $\zeta$ and the bulk relaxation time $\tau_\Pi$ of $npe$ matter (in the low temperature regime) depend on the symmetry energy and its experimentally measured parameters. Using a range of values for $S$ and $L$, we show that slight changes in $L$ can change the zero frequency limit of $\zeta$ by orders of magnitude. We also argue that the out-of-$\beta$-equilibrium correction to the pressure of $npe$ matter at saturation density present in the early inspiral phase of binary neutron star mergers dynamically responds to volume deformations in an approximate bulk elastic manner  \cite{Landau:1986aog}, with a bulk modulus determined by the values of $S$ and $L$.   We discuss the potential consequences of our findings for the detection of both conservative and dissipative effects encoded in the gravitational waves emitted by inspiralling neutron stars.


{\it Bulk viscosity from chemical imbalance} --- Bulk-viscous effects can be relevant to the damping of r-modes in neutron stars \cite{Madsen:1992sx, Kokkotas:2001ze, Reisenegger:2003pd,Weber:2004kj,Weber:2006iw, Alford:2010gw,Schmitt:2017efp} and possibly to the gravitational-wave emission in binary neutron star mergers \cite{Alford:2017rxf, Alford:2020pld,Most:2021zvc, Most:2022yhe, Hammond:2022uua, Espino:2023dei,Chabanov:2023blf,Chabanov:2023abq,CruzRojas:2024etx}. Furthermore, dissipative effects \cite{Ripley:2023qxo, HegadeKR:2024slr, Ripley:2023lsq} contribute to the so-called dissipative tidal deformability $\tilde\Xi$ \cite{Ripley:2023qxo}, which modifies the phase of gravitational waves emitted during the inspiral of a neutron star binary. An upper bound of $\tilde\Xi \lesssim 1200$ has been placed using data from the gravitational wave event GW170817 in Refs.\  \cite{Ripley:2023lsq,HegadeKR:2024slr}. Finally, we note that tidal heating has been proposed as a way to probe strangeness degrees of freedom in neutron stars \cite{Ghosh:2023vrx,Ghosh:2025glz}.

In this work we consider neutrino-transparent $npe$ matter in a macroscopic quasi-equilibrium state described by $\{s,n_B,Y, u^\mu\}$, where $s$ is the entropy density, $Y = n_e/n_B$ is the electron fraction (charge neutrality implies that the electron density, $n_e$, equals the proton density, $n_p$), and $u^\mu$ is the local 4-velocity (normalized such $u_\mu u^\mu = 1$). At sufficiently low temperatures, the $\beta-$equilibrium condition is $\mu_n = \mu_p + \mu_e $, where $\mu_n$, $\mu_p$, and  $\mu_e$ are the chemical potentials for neutrons, protons, and electrons, respectively. Therefore, in this  regime, deviations from $\beta-$equilibrium can be quantified by the nonzero value of the reaction affinity $\delta \mu \equiv \mu_n  - \mu_p - \mu_e$ (at high temperatures $\beta-$equilibrium does not imply $\delta\mu=0$, see \cite{Alford:2018lhf,Alford:2021ogv,Harris:2024ssp}). 

The dynamical equations that describe this system are \cite{Gavassino:2020kwo,Celora:2022nbp}
\begin{subequations}
\begin{align}
\nabla_{\mu}(s u^\mu) = \frac{Q+\delta \mu \,\Gamma_e}{T},\\
(\varepsilon+\mathcal{P})u^\mu \nabla_\mu u^\nu - \Delta^{\nu\alpha}\nabla_\alpha \mathcal{P} = 0,\\
\nabla_\mu ( n_B u^\mu)=0,\\
u^\mu \nabla_\mu Y = \frac{\Gamma_e}{n_B},
\end{align}
\end{subequations}
where $Q$ stems from the (assumed isotropic) radiative energy loss due to neutrinos and $\Gamma_e$ describes flavor equilibration rates associated with direct and modified Urca processes (see the Supplemental Material for details \cite{Yang_SM_2025}). Above, $T$ is the temperature, $\varepsilon = \varepsilon(s,n_B,Y)$ is the system's energy density, $\mathcal{P} = \mathcal{P}(s,n_B,Y)$ is the pressure,   
$\Delta_{\mu\nu} = g_{\mu\nu}-u_\mu u_\nu$, and $g_{\mu\nu}$ is the (mostly minus) spacetime metric.

For small deviations around $\beta-$equilibrium, one may approximate $\Gamma_e \sim \lambda\, \delta \mu$, where $\lambda \sim T^4$ for direct Urca and $\lambda \sim T^6$ for modified Urca processes (assuming, for simplicity, the so-called Fermi surface approximation, see \cite{Alford:2018lhf, Alford:2021ogv, Harris:2020rus} and the Supplemental Material \cite{Yang_SM_2025}). The out-of-$\beta-$equilibrium correction to the system's pressure, described by the bulk scalar $\Pi \equiv \mathcal{P}-\mathcal{P}|_{\delta \mu = 0}$ \cite{Gavassino:2020kwo}, obeys in this limit a dynamical equation  \cite{Gavassino:2020kwo, Camelio:2022ljs} \`a la Israel and Stewart \cite{Israel:1979wp}
\begin{equation}
\label{eq:ISfinalform}
    u^\mu \nabla_\mu \Pi = - \frac{\Pi}{\tau_\Pi} -\frac{\zeta_0}{\tau_\Pi} \nabla_\mu u^\mu ,
\end{equation}
where the transport coefficients are given by
\begin{subequations}
\label{eq:coefsthermo}
    \begin{align}
        \tau_{\Pi} &= - \frac{n_B}{\la} \left( \frac{\partial \delta \mu}{\partial Y}\bigg|_{n_B} \right)^{-1},\\
        \zeta_0 &= - \frac{P_1 \, n_B^2}{\la} \left( \frac{\partial \delta \mu}{\partial Y}\bigg|_{n_B} \right)^{-1} \left.\frac{\partial \delta \mu}{\partial n_B}\right|_{Y}. 
    \end{align}
\end{subequations}
and $P_1 = (\partial \mathcal{P}/\partial\delta\mu)_{n_B,\delta \mu = 0}$. Above, we neglected $T-$dependent effects on  the EOS. The contribution from $Q$ to \eqref{eq:ISfinalform} is also negligible, see the Supplemental Material \cite{Yang_SM_2025}. Finally, we note that while here we focus on the small $\delta \mu$ case, the equivalence between bulk-viscous theories and chemical equilibration dynamics holds in the full nonlinear regime, see \cite{Gavassino:2023xkt}. 

Equation \eqref{eq:ISfinalform} shows that small deviations in pressure away from $\beta-$equilibrium in $npe$ matter are governed by the equations of a relativistic viscoelastic material in which $\Pi$ displays a delay in response to time-dependent deformations \cite{Gavassino:2023xkt}. Assuming, for simplicity, small amplitude baryon density oscillations $\sim e^{-i\omega t}$ around a spatially uniform $\beta-$equilibrated state, linear response gives
\be\label{eq:linear_response}
\delta \Pi= - \frac{\zeta_0}{1-i\tau_\Pi \omega} \delta (\nabla_\mu u^\mu) = \frac{i}{\omega} G_R(\omega)\, \delta (\nabla_\mu u^\mu),
\ee
where we defined the retarded Green's function, $G_R$, that relates the stress and the expansion rate (see the Supplemental Material for details \cite{Yang_SM_2025}). The $\omega-$dependent bulk viscosity can be extracted from the imaginary part of the Green's function, 
\be
\zeta(\omega) = \frac{1}{\omega} \operatorname{Im} G_R = \frac{\zeta_0}{1 + \tau_\Pi^2 \omega^2},
\label{definezetaomega}
\ee
defined in \cite{Sawyer:1989dp} (for a review, see \cite{Harris:2024evy}). Different phenomena \footnote{We assume that $\sim 1/\omega$ and $\tau_\Pi$ are much larger than any microscopic equilibration time  $\tau_\mathrm{micro}$, see \cite{Gavassino:2020kwo}.} are encoded in \eqref{eq:ISfinalform}. When $\tau_{\Pi}\omega \ll 1$, one may find $\Pi \sim - \zeta_0 \nabla_\mu u^\mu$ and the system is in the Navier-Stokes viscous fluid regime \cite{Gavassino:2020kwo}. When $\tau_\Pi \omega \sim 1$, the system is in the resonant regime where $\Pi$ responds on the same timescale associated with macroscopic motion, leading to maximal dissipation. This regime may be realized in neutron star mergers, see \cite{Alford:2018lhf,Alford:2021ogv,Most:2021zvc,Most:2022yhe,Hammond:2022uua,Espino:2023dei}. On the other hand, when $\tau_\Pi \omega \gg 1$, the system is in the \emph{frozen} regime \cite{Haensel:2002qw,Gavassino:2020kwo,Montefusco:2024xrx} where $Y$ stays fixed, macroscopic motion becomes approximately reversible, and there is basically no dissipation \cite{Gavassino:2020kwo}.

Due to the strong temperature dependence of the Urca rates, $\tau_\Pi$ and $\zeta_0$ are strongly enhanced at low temperatures, and neutrino-transparent $npe$ matter is expected to be in the frozen $Y$ regime \cite{Haensel:2002qw,Gavassino:2020kwo,Montefusco:2024xrx}. 
In this case, because the equilibrium charge fraction $Y_{\textrm{eq}}$ depends on density while $Y$ remains constant, the system still drops out of $\beta-$equilibrium as matter compresses or expands. As in equilibrium, the full pressure becomes a function of the baryon density, but this function is given by the EOS at fixed proton fraction, which is not in $\beta-$equilibrium. This is relevant for early-to-late binary neutron star inspirals, where $T \sim 10^5 \mathrm{\; K}$, tidal forces deform the stars with frequencies $\om \sim  \mathrm{\; kHz}$ \cite{Ripley:2023lsq}, and $\delta \mu \neq 0$.

We here note that, formally, in this frozen regime $\zeta_0,\tau_\Pi \to \infty$,  $\zeta(\omega) \to \zeta_0/(\tau_\Pi \omega)^2$,  and the out-of-$\beta-$equilibrium pressure deviation $\Pi$ is approximately described by the following dynamical equation
\begin{equation}
\label{eq:ISelastic}
    u^\mu \nabla_\mu \Pi =  -\frac{\zeta_0}{\tau_\Pi} \nabla_\mu u^\mu ,
\end{equation}
which corresponds to a relativistic generalization of Hooke's law of elasticity \cite{Landau:1986aog,Gavassino:2023xkt}, even though of course there is no underlying lattice in this fluid. This approximate bulk elastic response is characterized by a bulk modulus coefficient given by $\zeta_0/\tau_\Pi$, which does not depend on the rates, see \eqref{eq:coefsthermo}, and is fully determined by the cold matter EOS. We also note that, in this limit, the negative bulk modulus is exactly the real part of $G_R$ defined in Eq.~\eqref{eq:linear_response} and represents local reactive conservative processes.


{\it Symmetry energy dependence of transport coefficients} --- 
The energy per baryon, $E(n_B, \delta)$, may be expanded in powers of the isospin asymmetry $\delta \equiv 1 - 2\,Y$ 
as follows 
\be
\bea
\label{SNM expansion}
E(n_B, \delta) &\approx E(n_B, \delta = 0) + E_{sym,2}(n_B)\de^2 \\
&\;\;\;\;+ E_{sym,4}(n_B)\de^4 + \mathcal{O}(\de^6),
\eea
\ee
where $E_{sym,2}(n_B)$, also denoted as $E_{sym}(n_B)$,  is usually referred to as the nuclear symmetry energy \cite{Li:1997px,Chen:2005ti,Li:2019xxz,Xu:2010fh,Li:2021thg}. Neglecting $\mathcal{O}(\delta^4)$ terms, \eqref{SNM expansion} is reduced to the empirical parabolic approximation \cite{PhysRevC.44.1892,Lattimer:1991ib}, and $E_{sym}(n_B)$ can be approximated by the difference between symmetric nuclear matter and pure neutron matter,
\be
E_{sym}(n_B) \approx E(n_B, \delta=1) - E(n_B, \delta = 0). 
\ee
This approximation has been shown to work reasonably well in the context of microscopic asymmetric matter calculations of chiral NN and 3N interactions at $n_B \lesssim n_{\mathrm{sat}}$ \cite{Drischler:2013iza, Drischler:2015eba, Kaiser:2015qia, Wellenhofer:2016lnl, Wellenhofer:2015qba} and in phenomenological models, even at higher densities, as shown for a chiral mean field model in a recent work by two of the authors and collaborators \cite{Yang:2025wop}. On the other hand, the expansion of $E_{sym}(n_B)$ around $n_\mathrm{sat}$ gives 
\be
\bea
E_{sym}(n_B) &= S + L \left( \frac{n_B - n_{\textrm{sat}}}{3n_{\textrm{sat}}} \right) \\
&\;\;\;\;+ \frac{K_{sym}}{2}\left( \frac{n_B - n_{\textrm{sat}}}{3n_{\textrm{sat}}} \right)^2 +\ldots, 
\eea
\ee
which defines $S$,  the symmetry slope $L$, the curvature $K_{sym}$, and other higher-order coefficients computed at saturation density.


We note here that the bulk viscosity can be connected to the symmetry energy expansion by noting that $\delta \mu$ is the derivative of the energy per baryon $E + E_{\ell}$ with respect to $n_I/n_B \equiv \delta/2$, where $E_{\ell}$ is the contribution from the leptons (treated here as a free electron gas). This gives
\be\label{eq:deltamu}
\de\mu = 4\, E_{sym}(n_B)(1-2Y) 
-
\left(\frac{\pa E_{\ell}}{\pa Y}\right)_{\substack{n_B}}.
\ee
Equation~\eqref{eq:deltamu} shows that $\delta \mu$ follows from the symmetry energy and the energy of the electrons. As a consequence, the EOS-part of the transport coefficients also depends solely on the symmetry energy and electron contributions to the energy per baryon. In fact, modulo the contribution from the rates and leptons, one finds schematically 
\be
\tau_\Pi \sim \frac{1}{E_{sym}(n_B)}, \;\;\; \zeta_0 \sim \left(\frac{\pa E_{sym}(n_B)}{\pa n_B}\bigg/E_{sym}(n_B)\right)^2.
\label{transport_coefficient_behavior}
\ee
Thus, as long as the parabolic approximation in the asymmetry parameter $\delta$ holds, the transport coefficients can be reliably calculated using experimentally observed properties of the symmetry energy \footnote{We note that Ref.\ \cite{Alford:2010gw} computed the bulk viscosity using a parabolic approximation of the APR  EOS \cite{Akmal:1998cf}.}. We remark that Eq.~\eqref{transport_coefficient_behavior} was derived without making an expansion in the baryon density. Nevertheless, we note that the quantities are only fully determined by $S$ and $L$ when they are evaluated at $n_{\mathrm{sat}}$. This shows that models for cold $npe$ matter can only give different values for $\zeta_0$ and $\tau_\Pi$ if they do not have the same basic nuclear symmetry properties (assuming they use similar rates).

\begin{figure}[!ht]
    \includegraphics[width=7.5cm]{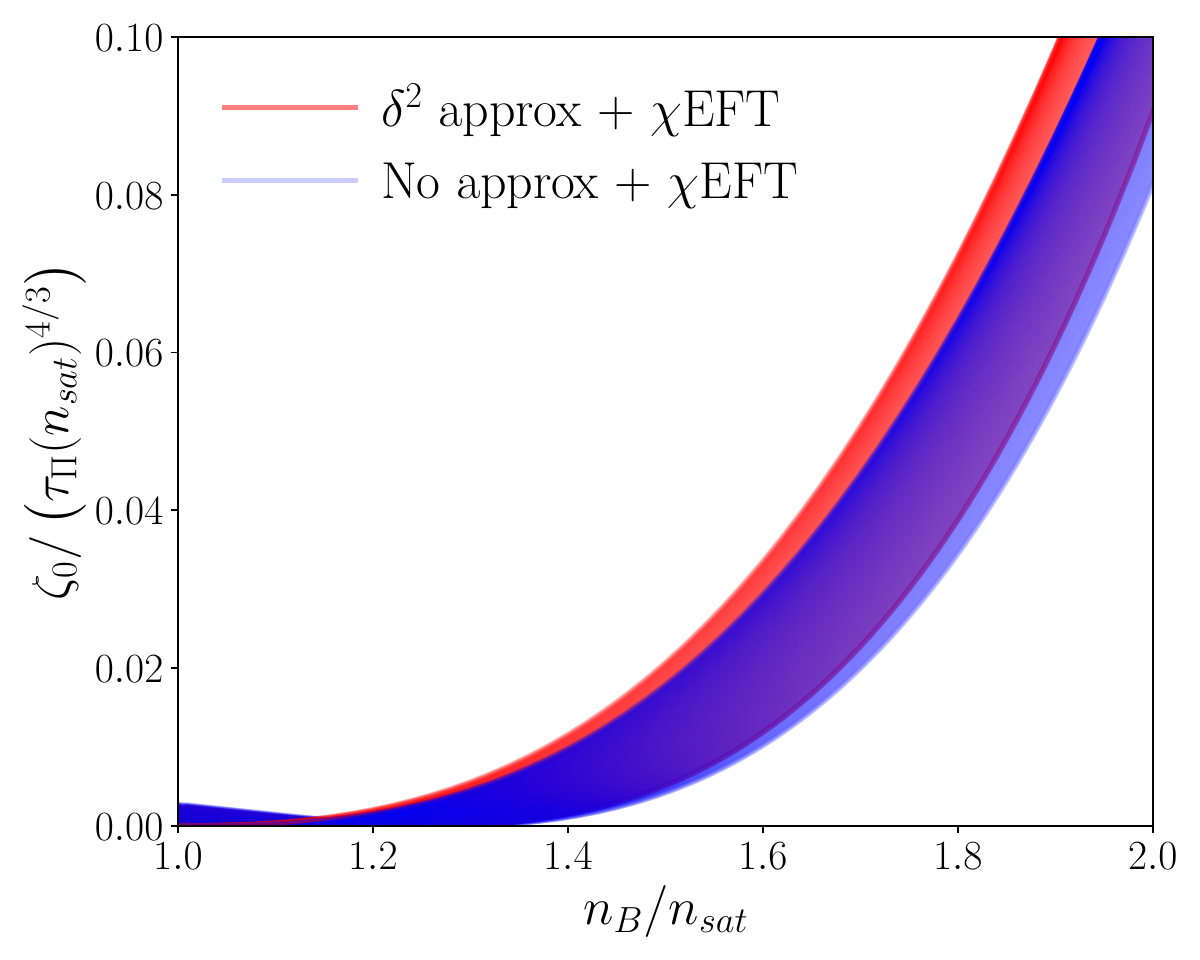}
    \caption{The dimensionless bulk modulus as a function of the baryon density. The blue curves are calculated directly from the chiral EFT parameterization \cite{Tews:2024owl, Hippert:2024hum, Hebeler:2010jx, Bedaque:2014sqa, Tews:2018kmu}. The red curves are calculated using the parabolic symmetry energy approximation of the chiral EFT parameterization. The purple region denotes the overlap between blue and red curves (darker colors reflect a higher density of curves).}
    \label{plot:bulk_modulus_chiral}
\end{figure}

\begin{figure}[!ht]
    \includegraphics[width=7.5cm]{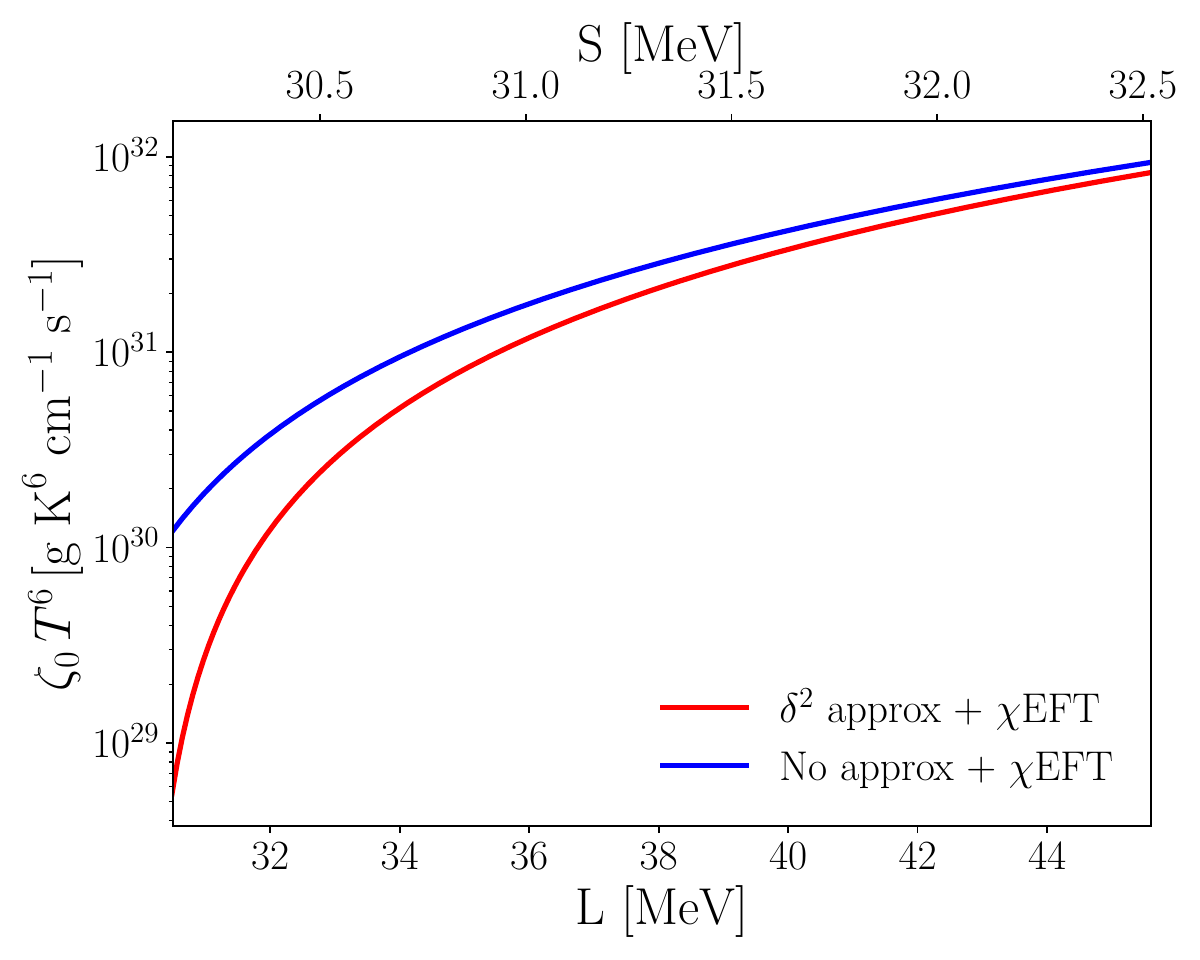}
    \caption{Bulk viscosity at $n_{\mathrm{sat}}$ as a function of $S$ or $L$. The blue curve is calculated directly from a chiral EFT parameterization \cite{Tews:2024owl, Hippert:2024hum, Hebeler:2010jx, Bedaque:2014sqa, Tews:2018kmu}. The red curve is calculated using the parabolic symmetry energy approximation of the chiral EFT parameterization.}
    \label{plot:SL_correlation}
\end{figure}

To assess the validity of the parabolic approximation in the calculation of bulk-viscous transport coefficients beyond RMF modeling, we calculate $\zeta_0$ and $\tau_\Pi$ using as input chiral EFT calculations from Ref.~\cite{Tews:2024owl}, which provided uncertainty bands for three different nuclear Hamiltonians. Following \cite{Hippert:2024hum}, we cover the entire band of energy per nucleon $E$ by linearly interpolating between the upper and lower bounds of chiral EFT, $E_{\mathrm{up}}$ and $E_{\mathrm{low}}$, 
\be
E_{\si} = (1 - \si)E_{\mathrm{up}} + \si E_{\mathrm{low}}. 
\ee
To perform the interpolation, we use the following parametrization \cite{Hebeler:2010jx, Bedaque:2014sqa, Tews:2018kmu}, 
\be\label{eq:chiralEFT}
\bea
E(n_B, Y) &= m_B   \\
&\;\; + T_0 \left[ \frac{3}{5}\left( Y^{3/5} + (1-Y)^{3/5} \right) \left(\frac{2n_B}{n_{\mathrm{sat}}}\right)^{2/3} \right.\\
&\;\; \left.- \left[ (2\al - 4\al_L)Y(1-Y) + \al_L \right]\frac{n_B}{n_{\mathrm{sat}}} \right.\\
&\;\; \left. + \left[(2\eta - 4\eta_L)Y(1-Y) + \al_L\right] \left(\frac{2n_B}{n_{\mathrm{sat}}}\right)^{\ga} \right], 
\eea
\ee
where $\al$, $\al_L$, $\eta$, $\eta_L$, and $\ga$ are fitting parameters \footnote{The parameters used in this paper can be found in \cite{yang_2025_17555031}.}. After fitting \eqref{eq:chiralEFT}, we calculate $S$, $L$, $\tau_\Pi$ and $\zeta_0$ for each obtained EOS (assuming $n_\mathrm{sat}=0.16$ fm$^{-3}$). In Fig.~\ref{plot:bulk_modulus_chiral}, we plot the dimensionless bulk modulus $\zeta_0/(\tau_\Pi n_\mathrm{sat}^{4/3})$ as a function of $n_B$ for all chiral EFT parameterizations. In Fig.~\ref{plot:SL_correlation}, we plot $\zeta_0$ at $n_{\mathrm{sat}}$ as a function of $S$ or $L$, which also shows the correlation between $S$ and $L$ for our chiral EFT parameterizations. The blue curves are calculated directly from the chiral EFT parametrization, while the red curves are calculated using the respective parabolic symmetry energy approximations of the chiral EFT parameterizations.

One can see that the parabolic approximation in the asymmetry parameter $\delta$ is excellent, especially at low densities and relatively higher $L$. This indicates that \eqref{transport_coefficient_behavior} very accurately captures the dependence of $\tau_\Pi$ and $\zeta_0$ with $E_{sym}$ in neutrino-transparent $npe$ matter. Also, even with the tight constraints on $L$ imposed from chiral EFT \cite{Tews:2024owl}, $\zeta_0$ remains strongly sensitive and changes by orders of magnitude when $L$ changes.

{\it Transport coefficients at saturation} --- 
The EOS-dependent contributions to the transport coefficients significantly simplify at saturation density, being solely determined by the experimentally measured quantities $S$ and $L$. In fact, direct calculation \footnote{Similar results were independently obtained by S.~P.~Harris (private communication).} gives
\begin{subequations}
    \begin{align}
        \label{eq:tauPi} \tau_\Pi(n_{\mathrm{sat}}) &= \frac{n_{\mathrm{sat}}}{\la} \left[ 8S + \frac{\pa^2 E_\ell}{\pa Y^2}\right]^{-1}, \\
        \label{eq:zeta}\zeta_0(n_{\mathrm{sat}}) &= \frac{n_{\mathrm{sat}}^4}{\la} \left[ 8S + \frac{\pa^2 E_\ell}{\pa Y^2}\right]^{-2} \nonumber\\
        &\;\;\;\; \times \left[ \frac{4L}{3n_{\mathrm{sat}}}(2Y-1) + \frac{\pa^2 E_\ell}{\pa n_B\pa Y} - \frac{1}{n_{\mathrm{sat}}}\frac{\pa E_\ell}{\pa Y} \right]^2. 
    \end{align}
\end{subequations}

\begin{figure}[!ht]
    \includegraphics[width=7.5cm]{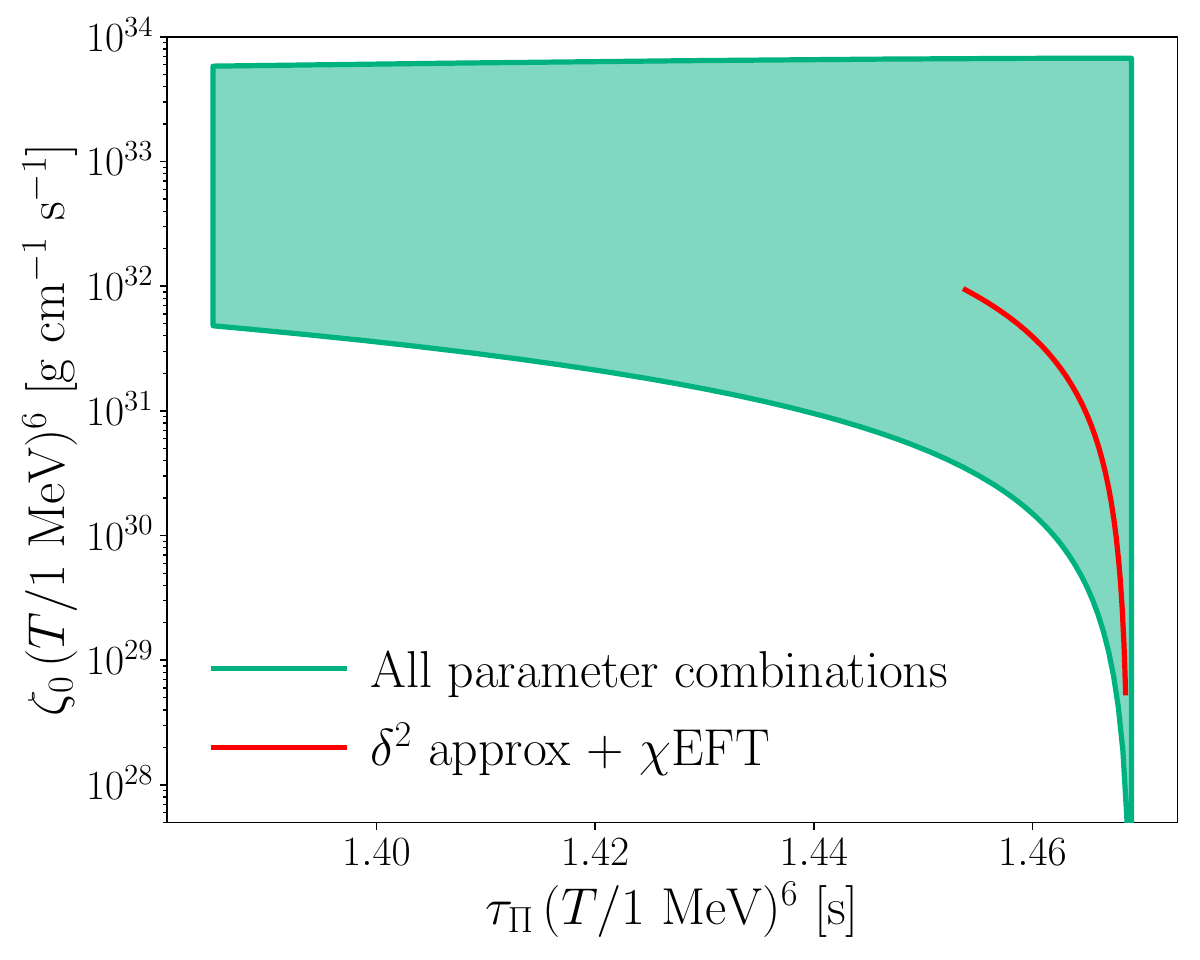}
    \caption{Bulk viscosity vs. relaxation time at $n_{\mathrm{sat}}$. The green box is calculated using $S \in [30, 40]$ MeV and $L \in [30, 150]$ MeV. The red curve is calculated using the chiral EFT parameterizations \cite{Tews:2024owl, Hippert:2024hum, Hebeler:2010jx, Bedaque:2014sqa, Tews:2018kmu}.}
    \label{plot:zeta}
\end{figure}

In Fig.~\ref{plot:zeta}, we take $S \in [30, 40]$ MeV and $L \in [30, 150]$ MeV, and we plot $\zeta_0$ vs. $\tau_\Pi$ using all combinations of $S$ and $L$, without taking into account their correlations. As the figure shows, $\zeta_0(n_\mathrm{sat})$ is strongly sensitive to $L$.

Using \eqref{eq:tauPi} and \eqref{eq:zeta}, one can see that the nuclear physics contribution to the bulk modulus $\zeta_0/\tau_\Pi$ is fully determined by $n_\mathrm{sat}$, $S$, and $L$ (above $n_\mathrm{sat}$, the bulk modulus depends on higher order coefficients such as $K_{sym}$). We find $\zeta_0/(\tau_\Pi n_\mathrm{sat}^{4/3}) = 0.004$ for $L\sim 50$ MeV and $\zeta_0/(\tau_\Pi n_\mathrm{sat}^{4/3}) = 0.056$ for $L\sim 100$ MeV. Tighter constraints on $L$ are needed to nail down the value of the bulk modulus of $npe$ matter. 

{\it Final remarks} --- 
In this letter, we have shown that the bulk-viscous transport properties $\zeta_0$ and $\tau_\Pi$ of $npe$ matter at saturation density are directly constrained by knowledge on the coefficients $S$ and $L$ of the symmetry energy expansion. 
In particular, $\zeta_0$ can vary by orders of magnitude by varying $L$. 
The strong dependence of $\zeta_0$ with $L$ is particularly relevant given the upper bound on the dissipative tidal deformability found in Ref.\ \cite{Ripley:2023qxo}, which was translated into an upper bound on a density-averaged bulk viscosity $\langle \zeta \rangle_A\lesssim 10^{31}$ g cm$^{-1}$$s^{-1}$ in the inspiralling neutron stars \cite{Ripley:2023qxo}. One can see that $\zeta_0$ at saturation can become much larger than this bound on $\langle \zeta \rangle_A$ if $L$ increases. In addition, as illustrated by Eq.~\eqref{eq:ISelastic}, the Israel-Stewart equation also captures conservative dynamics and should, thus, contribute to the \emph{conservative} dynamical Love number $k_l^{(2)}$ \cite{HegadeKR:2024agt}. As Fig.~\ref{plot:zeta} shows, the bulk modulus $\zeta_0/\tau_\Pi$ can increase by orders of magnitude as $L$ increases, and is not suppressed by temperature. Therefore, we also expect a potentially detectable effect induced by bulk viscosity through the conservative dynamical tides, though the precise relation requires a new dedicated analysis that is left for future work. These findings suggest the interesting possibility that one may use both the conservative and dissipative tidal response to place an upper bound on $L$. While our results are limited to nuclear saturation and slightly larger densities, they should be informative of the averaged bulk viscosity computed taking into account the different layers of the star.

For the dissipative processes, it is unclear how the upper bound on the bulk viscosity from \cite{Ripley:2023qxo} is affected by the very large relaxation times found at low temperatures, which induce approximate elastic response of the bulk scalar $\Pi$. Furthermore, it is important to remark that our results are only valid for the transport coefficients of $npe$ matter at low temperatures and at saturation density. The average $\langle \zeta \rangle_A$ should also receive contributions from the stellar crust, below $n_{\textrm{sat}}$, and from the inner core, where $n_B$ can reach a few times $n_{\textrm{sat}}$ and exotic degrees of freedom, such as hyperons and quarks, may be found and contribute to bulk viscosity. Using results from Ref. \cite{Yakovlev:2018jia}, we estimate that densities below saturation at the stellar crust should contribute less than 1\% to $\langle \zeta \rangle_A$. Nevertheless, the precise contributions to $\langle \zeta \rangle_A$ from the inner core remain to be determined.  
Further studies, combining general-relativistic calculations with a more complete description of neutron star matter from crust to core, 
are needed to determine how gravitational waves emitted during the early-to-late inspiral may encode information on the nuclear symmetry energy.


{\it Acknowledgments} ---
We thank I.~Tews for providing the chiral EFT data, L.~Gavassino, S.~P.~Harris, A.~Hegade, E. R.~Most, and L.~Rezzolla for enlightening discussions. 
JN and YY are partly supported by the U.S. Department of Energy, Office of Science, Office for Nuclear Physics
under Award No. DE-SC0023861. 
M.H. was supported in part by the Universidade Estadual do Rio de Janeiro through the Programa de Apoio à Docência (PAPD), and by the Brazilian National Council for Scientific and Technological Development (CNPq) under process No. 313638/2025-0.
E.S. has received funding from the European Union’s Horizon Europe research and innovation program under the Marie Sk\l odowska-Curie grant agreement No. 101109747.

\bibliography{references,notInspire}

\clearpage

\onecolumngrid

\section{Supplemental material}
In this Supplementary Material, we provide details about the reaction rates, the neutrino emissivity, the derivation of the bulk-viscous transport coefficients for both the neutrino-trapped and neutrino-transparent regimes, and their dependence on the nuclear symmetry energy.

\subsection{Urca rates}
\label{A: rate}
Within the Fermi surface approximation, one can obtain analytical expressions for the Urca rates \cite{Alford:2018lhf, Alford:2021ogv, Harris:2020rus} as a sum of individual rates, 
\be
\Ga_e = \Ga_{dU} + \Ga_{mU,n} + \Ga_{mU,p},
\ee
where
\begin{subequations}
    \begin{align}
        &\Ga_{dU} = \frac{G^2(1+3g_A^2)}{240\pi^5}E^*_{Fn}E^*_{Fp}p_{Fe}\vartheta_{dU}\de\mu(17 \pi^4T^4 + 10\pi^2\de\mu^2T^2 + \de\mu^4) , \\
        &\Ga_{mU,n} = \frac{1}{5760\pi^9}G^2 g_A^2 f^4 \frac{(E^*_{Fn})^3 E^*_{Fp}}{m_\pi^4} \frac{p_{Fn}^4 p_{Fp}}{(p_{Fn}^2+m_\pi^2)^2}\vartheta_n \de\mu( 1835 \pi^6 T^6 + 945\pi^4 \de\mu^2 T^4 + 105\pi^2\de\mu^4 T^2 + 3\de\mu^6 ) ,\\
        &\Ga_{mU,p} = \frac{1}{40320\pi^9}G^2 g_A^2 f^4 \frac{E^*_{Fn} (E^*_{Fp})^3}{m_\pi^4} \frac{p_{Fn}(p_{Fn}-p_{Fp})^4}{((p_{Fn}-p_{Fp})^2+m_\pi^2)^2}\vartheta_p \de\mu ( 1835\pi^6 T^6 + 945\pi^4 \de\mu^2 T^4 + 105\pi^2\de\mu^4T^2 + 3\de\mu^6 ).
    \end{align}
\end{subequations}
Above, the pion-nucleon coupling constant is $f\approx1$, $G^2 = G_F^2 \,cos^2\theta_c = 1.1\times 10^{-22} $MeV$^{-4}$, $G_F$ is the Fermi coupling constant, $\theta_c$ is the Cabibbo angle, the axial vector coupling constant is $g_A = 1.26$, $p_{FN}$ is the nucleon Fermi momentum, and $E^*_{FN} = \sqrt{p_{FN}^2 + m_N^{*2}}$ is the nucleon energy. 

In the Fermi surface approximation, the direct Urca rate is only turned on if the triangle relation is satisfied,
\be
  \vartheta_{dU} =
  \begin{cases}
  1 & \text{if $p_{Fn}<p_{Fp} + p_{Fe}$} \\
  0 & \text{if $p_{Fn}>p_{Fp} + p_{Fe}$}.
  \end{cases}
\ee
The modified Urca rate is also modified by the density through,
\begin{subequations}
    \begin{align}
        \vartheta_n &=
        \begin{cases}
        1 & \text{if $p_{Fn}>p_{Fp} + p_{Fe}$} \\
        1 - \frac{3}{8} \frac{(p_{Fp} + p_{Fe} - p_{Fn})^2}{p_{Fp}p_{Fe}} & \text{if $p_{Fn}<p_{Fp} + p_{Fe}$} ,
        \end{cases} \\
        \vartheta_p &=
        \begin{cases}
        0 & \text{if $p_{Fn}>3p_{Fp} + p_{Fe}$} \\
        \frac{(3p_{Fp} + p_{Fe} - p_{Fn})^2}{p_{Fn}p_{Fe}} & \text{if} 
        \begin{cases}
        p_{Fn}>3p_{Fp} - p_{Fe} \\
        p_{Fn}<3p_{Fp} + p_{Fe}
        \end{cases} \\
        4 \frac{3p_{Fp} - p_{Fn}}{p_{Fn}} &
        \text{if}
        \begin{cases}
        3p_{Fp} - p_{Fe} > p_{Fn} \\
        p_{Fn} > p_{Fp} + p_{Fe}
        \end{cases} \\
        2 + 3\frac{2p_{Fp} - p_{Fn}}{p_{Fe}} - 3\frac{(p_{Fp} - p_{Fe})^2}{p_{Fn}p_{Fe}} & \text{if $p_{Fn}<p_{Fp} + p_{Fe}$} .
        \end{cases}
    \end{align}
\end{subequations}

In this work, we only keep $\Ga_e$ up to linear order in $\delta\mu$, which defines the coefficient $\lambda$ via $\Ga_e \approx \la \de\mu$.


\newpage
\subsection{Neutrino emissivity}

Local conservation of energy and momentum is described by
\be
\nabla_\mu T^{\mu\nu} = Q^\nu,
\ee
where $Q^\nu$ describes the radiative loss due to neutrinos. If neutrinos are trapped, then $Q^\nu = 0$. If the neutrinos are not trapped, assuming that the emission is isotropic in the fluid rest frame, we can write
\be
Q^\nu = - Q u^\nu, 
\ee
where $Q$ is the total luminosity. In the limit of small $\de\mu$, one can obtain the approximate expression \cite{Camelio:2022ljs},
\be
Q = \frac{1.22 \times 10^{25}}{\text{erg}^{-1} \text{cm}^3 \text{s}}\left(\frac{n_e}{n_n}\right)^{1/3} \left( \frac{T}{10^9\text{K}} \right)^6 \frac{1}{60}\left[ 
\frac{457}{21} \pi^6 + 51\pi^4 \left(\frac{\de\mu}{T}\right)^2 + 15\pi^2 \left(\frac{\de\mu}{T}\right)^4 + \left(\frac{\de\mu}{T}\right)^6 \right]. 
\ee
Thus, $Q \sim Q_0 + \mathcal{O}(\delta \mu^2)$, with $Q_0 \sim T^6$. The quadratic dependence on $\delta\mu$ shows that, at sufficiently low temperatures, one may neglect the effect of $Q$ in the effective Israel-Stewart like dynamical equations for the bulk scalar $\Pi$.


\subsection{Entropy production}

Defining the entropy current as
\begin{equation}
    s^\mu = s u^\mu ,
\end{equation}
where $s$ is the entropy density, using the first law of thermodynamics, the entropy production rate is determined to be
\begin{equation}
    \begin{aligned}
    \nabla_\mu s^\mu & = u^\mu \nabla_\mu s + s \theta \\
    &= \frac{1}{T} \left( \delta \mu \,\Gamma_e + Q \right),
    \end{aligned}
\end{equation}
where $\theta\equiv \nabla_\mu u^\mu$ is the expansion rate scalar.


\newpage
\subsection{Israel-Stewart formalism in the linear $\de \mu$ regime}
\label{A: linear}

For convenience, here we provide the derivation of the Israel-Stewart equation for the bulk scalar used in the main text. We follow the derivation presented in \cite{Yang:2023ogo}, which was based on the original reference \cite{Gavassino:2020kwo}.

Assuming a small linear perturbation in terms of $\de\mu$, we can expand the pressure around $\beta-$equilibrium,
\be
\mathcal{P}(s, n_B, Y) = \mathcal{P}|_{\delta \mu = 0} + \Pi.
\ee
where we define the bulk scalar $\Pi = P_1 \delta \mu$ and $P_1 = \left.\frac{\partial \mathcal{P}}{\partial \delta \mu}\right|_{s, n_B, \delta \mu = 0}$. The system can be fully described by entropy production, baryon number conservation, and a non-conserved charge current characterized by the reaction rate,
\be
\label{conservation}
\nabla_\mu s^\mu = \frac{Q}{T} + \frac{\delta \mu \,\Gamma_e}{T} ,\;\;\;
\nabla_\mu (n_B u^\mu) = 0,\;\;\;
u^\mu \nabla_\mu Y = \frac{\Gamma_e}{n_B}, 
\ee
If we define partial equilibrium states with $\{s,n_B,Y\}$, we can write the change in $\de\mu$ as
\begin{equation}
\label{beta chain}
    u^\mu \nabla_\mu \delta \mu = \frac{\partial \delta \mu}{\partial s}\bigg|_{n_B, Y} u^\mu \nabla_\mu s + \frac{\partial \delta \mu}{\partial n_B}\bigg|_{s, Y} u^\mu \nabla_\mu n_B + \frac{\partial \delta \mu}{\partial Y}\bigg|_{s, n_B} u^\mu \nabla_\mu Y.
\end{equation}
Substituting Eq.~\eqref{conservation} into Eq.~\eqref{beta chain}, we find
\be
u^\mu \nabla_\mu \delta \mu = \frac{\partial \delta \mu}{\partial Y}\bigg|_{s, n_B} \frac{\la}{n_B} \de\mu + \frac{\partial \delta \mu}{\partial s}\bigg|_{n_B, Y} \frac{Q_0}{T} - \nabla_\mu u^\mu\, \left( n_B \left.\frac{\partial \delta \mu}{\partial n_B}\right|_{s, Y} + s \frac{\partial \delta \mu}{\partial s}\bigg|_{n_B, Y} \right),
\ee
where we keep terms up to linear order in  $\nabla_\mu s^\mu \approx {Q_0}/{T}$ and $n_B u^\mu\nabla_\mu Y = \Ga_e \approx \la \de\mu$. 
We can multiply this equation by $P_1$ and add $\de\mu\, u^\mu \nabla_\mu P_1$ to obtain
\be
\label{delta mu eom}
u^\mu \nabla_\mu \Pi = \frac{\partial \delta \mu}{\partial Y}\bigg|_{s, n_B} \frac{\la}{n_B} \Pi + \frac{\partial \delta \mu}{\partial s}\bigg|_{n_B, Y} \frac{Q_0}{T}P_1 - \nabla_\mu u^\mu \, P_1 \left( n_B \left.\frac{\partial \delta \mu}{\partial n_B}\right|_{s, Y} + s \frac{\partial \delta \mu}{\partial s}\bigg|_{n_B, Y} \right) + \de\mu\, u^\mu \nabla_\mu P_1. 
\ee
We can also consider the change $P_1$ in terms of all the non-conserved variables. Because $P_1$ is determined at $\delta \mu = 0$, we can write
\begin{equation}
    P_1 = P_1(s, n_B).
\end{equation}
Then
\begin{equation}
    u^\mu \nabla_\mu P_1 = \frac{\partial P_1}{\partial s} \bigg|_{n_B} u^\mu \nabla_\mu s + \frac{\partial P_1}{\partial n_B}\bigg|_s u^\mu \nabla_\mu n_B.
\end{equation}
Substituting Eq.~\eqref{conservation} into the equation above while keeping leading-order terms, we find
\begin{equation}
\label{P1 expansion}
    u^\mu \nabla_\mu P_1 =\frac{\partial P_1}{\partial s} \bigg|_{n_B} \left( \frac
    {Q_0}{T} - s \nabla_\mu u^\mu \right) - \frac{\partial P_1}{\partial n_B}\bigg|_s n_B \nabla_\mu u^\mu.
\end{equation}
Finally, substituting Eq.~\eqref{P1 expansion} into Eq.~\eqref{delta mu eom} gives
\be
\label{eom pre}
u^\mu \nabla_\mu \Pi = \frac{\partial \delta \mu}{\partial Y}\bigg|_{s, n_B} \frac{\la}{n_B} \Pi + \frac{\partial \Pi}{\partial s}\bigg|_{n_B, Y} \frac{Q_0}{T} - \nabla_\mu u^\mu \, P_1 \left( n_B \left.\frac{\partial \delta \mu}{\partial n_B}\right|_{s, Y} + s \frac{\partial \delta \mu}{\partial s}\bigg|_{n_B, Y} \right) - \nabla_\mu u^\mu \, \de\mu \left( n_B \left.\frac{\partial P_1}{\partial n_B}\right|_{s, Y} + s \frac{\partial P_1}{\partial s}\bigg|_{n_B, Y} \right).
\ee

If we define
\begin{subequations}
    \begin{align}
        &\tau_{\Pi} = - \frac{n_B}{\la} \left( \frac{\partial \delta \mu}{\partial Y}\bigg|_{s, n_B} \right)^{-1},\\
        &\zeta_0 = - \frac{n_B}{\la} \left( \frac{\partial \delta \mu}{\partial Y}\bigg|_{s, n_B} \right)^{-1} P_1 \left( n_B \left.\frac{\partial \delta \mu}{\partial n_B}\right|_{s, Y} + s \frac{\partial \delta \mu}{\partial s}\bigg|_{n_B, Y} \right), \\
        &\de_{\Pi\Pi} = - \frac{n_B}{\la P_1} \left( \frac{\partial \delta \mu}{\partial Y}\bigg|_{s, n_B} \right)^{-1} \left( n_B \left.\frac{\partial P_1}{\partial n_B}\right|_{s, Y} + s \frac{\partial P_1}{\partial s}\bigg|_{n_B, Y} \right),
        \end{align}
\end{subequations}
one can see that Eq.~\eqref{eom pre} becomes 
\begin{equation}
\label{eq:ISfinalform}
    u^\mu \nabla_\mu \Pi = - \frac{1}{\tau_\Pi} \left( \frac{\partial \Pi}{\partial s}\bigg|_{n_B, Y} \frac{Q_0}{T} + \Pi\right) - \frac{\zeta_0}{\tau_\Pi} \nabla_\mu u^\mu - \frac{\de_{\Pi\Pi} \Pi}{\tau_\Pi}\nabla_\mu u^\mu, 
\end{equation}
which is the Israel-Stewart equation with coefficients computed in $\beta-$equilibrium \cite{Gavassino:2020kwo}. Above, $\tau_\Pi$ is the bulk relaxation time, $\zeta_0$ is the bulk viscosity, and $\de_{\Pi\Pi}$ is a second-order hydrodynamic coefficient. In particular, the bulk modulus $\zeta_0/\tau_\pi$ is independent of the rates,
\be
\frac{\zeta_0}{\tau_\Pi} = P_1 \left( n_B \left.\frac{\partial \delta \mu}{\partial n_B}\right|_{s, Y} + s \frac{\partial \delta \mu}{\partial s}\bigg|_{n_B, Y} \right).
\ee

If we only keep the leading-order transport coefficients, Eq.~\eqref{eq:ISfinalform} becomes
\be
\label{eq:IS1storder}
u^\mu \nabla_\mu \Pi = - \frac{1}{\tau_\Pi} \left( \frac{\partial \Pi}{\partial s}\bigg|_{n_B, Y} \frac{Q_0}{T} + \Pi\right) - \frac{\zeta_0}{\tau_\Pi} \nabla_\mu u^\mu .
\ee

At sufficiently low temperatures, where we can use the parabolic approximation for the symmetry energy, we drop $Q_0$ and the equation becomes the one used in the main text
\begin{equation}
\label{IS:T0}
    u^\mu \nabla_\mu \Pi = - \frac{\Pi}{\tau_\Pi} - \frac{\zeta_0}{\tau_\Pi} \nabla_\mu u^\mu, 
\end{equation}
with the transport coefficients,
\begin{subequations}
\begin{align}
    &\tau_{\Pi} = - \frac{n_B}{\la} \left( \frac{\partial \delta \mu}{\partial Y}\bigg|_{n_B} \right)^{-1},\\
    &\zeta_0 = - \frac{P_1 \, n_B^2}{\la} \left( \frac{\partial \delta \mu}{\partial Y}\bigg|_{n_B} \right)^{-1} \left.\frac{\partial \delta \mu}{\partial n_B}\right|_{Y}. 
\end{align}
\end{subequations}
In this case, one can also see that the bulk modulus $\zeta_0/\tau_\Pi$ is given by
\be
\frac{\zeta_0}{\tau_\Pi} = P_1 n_B  \left.\frac{\partial \delta \mu}{\partial n_B}\right|_{Y}. 
\ee


\newpage
\subsection{Details on how the transport coefficients depend on the symmetry energy}

Without making any assumptions about the equation of state, one can always write the energy density of the system as a function of entropy density $s$, $n_B$, and the proton fraction $Y \equiv n_p/n_B$,
\be
\ep(s, n_B, Y) = m_B n_B f(s, n_B, Y), 
\ee
where $m_B$ is the vacuum baryon mass and $f(s, n_B,Y)$ is a dimensionless function.

We can further rewrite $\ep$ as a sum of a contribution from the symmetric nuclear matter and a contribution from the asymmetric part,
\be
\label{general sum}
\ep(s, n_B, Y) = m_B n_B \left[ f_1 \left(s, n_B\right) + f_2 \left(s, n_B,Y\right) \right],
\ee
where we define
\begin{subequations}
    \begin{align}
        &f_1\left(n_B\right) = f(s, n_B, Y = 1/2), \\
        &f_2 \left(s, n_B, Y\right) = f\left(s, n_B, Y\right) - f(s, n_B, Y = 1/2).
    \end{align}
\end{subequations}
At low temperatures, we can adopt the parabolic approximation for the symmetry energy and conclude that 
\be
m_B f_2(n_B,Y) = E_{sym}(n_B)(2Y - 1)^2. 
\ee
To describe $npe$ matter, we also need to add the electron contribution, so we have
\be
\label{Esym match}
m_B f_2(n_B,Y) = E_{sym}(n_B)(2Y - 1)^2 + E_l, 
\ee
where $E_l$ is the energy per baryon for electrons (we neglect the muon contribution in this work). We also note that at low temperatures where the neutrino mean free path is larger than the star radius \cite{Alford:2018lhf}, the neutrino contribution can be ignored, i.e. $E_{\nu_e} = 0$.

Without loss of generality, we can write the first law of thermodynamics for the system,
\be 
\label{first law}
d\ep = \frac{\ep + P}{n_B}dn_B - \de\mu\,n_B\,dY
\ee
and, using \eqref{general sum}, \eqref{Esym match} and \eqref{first law}, one finds,
\be
\de\mu = -\left( 4 E_{sym}(n_B)(2Y - 1) +  \left.\frac{\pa E_l}{\pa Y}\right|_{n_B} \right).
\ee
One can see that the out-of-$\beta-$equilibrium effect can be completely described by the symmetry energy and the electron contribution.  At $T=0$, the electrons can be described as free Fermi gas with
\be
E_l = \frac{3}{n_B \pi^2}\int^{(3\pi^2 n_B Y)^{1/3}}_0 dk\,k^2\sqrt{k^2 + m_e^2},
\ee
Similar to $\de\mu$, the transport coefficients can be calculated. Specifically, $\tau_\Pi$ and $\zeta_0$ are
\begin{subequations}
    \begin{align}
        \tau_\Pi &= \frac{n_B}{\la} \left( 8 E_{sym}(n_B) + \left.\frac{\pa^2 E_l}{\pa Y^2}\right|_{n_B}\right)^{-1}, \\
        \zeta_0 &= \frac{n_B^4}{\la} \left( 8 E_{sym}(n_B) + \frac{\pa^2 E_l}{\pa Y^2}\right)^{-2} \\
        &\;\;\;\; \times \left[ 4 \frac{\pa E_{sym}(n_B)}{\pa n_B}(2Y - 1) + \left( \frac{\pa^2 E_l}{\pa n_B\pa Y} - \frac{1}{n_B}\frac{\pa E_l}{\pa Y} \right) \right]^2. 
    \end{align}
\end{subequations}
Thus, as long as the parabolic approximation holds, the transport coefficients can be calculated using experimental information about the symmetry energy. Also, we can directly conclude that the relaxation time increases with $E_{sym}$, and the bulk viscosity increases with the derivative of $E_{sym}$, as discussed in the main text.

Furthermore, at saturation density, $\tau_\Pi$ and $\zeta_0$ reduce to
\begin{subequations}\label{eq:transport_nsat}
    \begin{align}
        \tau_\Pi(n_{\mathrm{sat}}) &= \frac{n_{\mathrm{sat}}}{\la} \left[ 8S + \frac{\pa^2 E_l}{\pa Y^2}\right]^{-1},  \\
        \zeta_0(n_{\mathrm{sat}}) &= \frac{n_{\mathrm{sat}}^4}{\la} \left[ 8S +  \frac{\pa^2 E_l}{\pa Y^2}\right]^{-2} \\
        &\;\;\;\; \times \left[ 4 \frac{L}{3n_{\mathrm{sat}}}(2Y - 1) + \frac{\pa^2 E_l}{\pa n_B\pa Y} - \frac{1}{n_{\mathrm{sat}}}\frac{\pa E_l}{\pa Y} \right]^2. 
    \end{align}
\end{subequations}
We can immediately see that these transport coefficients can be directly computed from the experimentally constrained quantities $S$ and $L$.


\newpage
\subsection{Linear response theory for the Israel-Stewart equation}
\label{A: AC IS}

One can derive a retarded Green's function from the Israel-Stewart equation and study both the conservative and dissipative effects, assuming a linear perturbation of the spacetime metric, $\de g^{\mu\nu} \propto e^{i\om t}$. The metric perturbation induces the expansion rate,
\begin{equation}
   \nabla_\mu u^\mu = \frac{ \partial_0 \sqrt{-\det(\eta_{\mu\nu} + \delta g_{\mu\nu})}}{\sqrt{-\det(\eta_{\mu\nu})}} = \partial_0\sqrt{1 - 2\,\eta^{\alpha\beta}\delta g_{\alpha\beta}}= -i\omega\,\eta^{\alpha\beta}\delta g_{\alpha\beta}.
\end{equation}

One can substitute this expansion rate into Eq.~\eqref{eq:IS1storder} to get, 
\begin{equation}
    -i\omega \,\tau_{\Pi}\, \Pi + \Pi = i\omega\,\zeta_0\,\eta^{\alpha\beta}\delta g_{\alpha\beta}.
\end{equation}
Because $\Pi$ is the correction to the equilibrium pressure, $\Pi=\delta p = \delta T^i_i/3$. The equation becomes
\begin{equation}
\label{eq:linearresp}
    \Pi = \frac{1}{3}\delta T^i_i = \frac{i\omega\, \zeta_0}{1-i\omega\,\tau_{\Pi}}\eta^{\alpha\beta}\delta g_{\alpha\beta}. 
\end{equation}
One can then extract the retarded Green's function,
\begin{equation}
    G^{\theta}_R(\omega) = \frac{i\omega\,\zeta_0}{1-i\omega \tau_{\Pi}} = \operatorname{Re}G^{\theta}_R(\omega) + i\operatorname{Im}G^{\theta}_R(\omega)\,,
\end{equation}
where we separate the real and imaginary parts,
\begin{equation}
     \operatorname{Re}G^{\theta}_R(\omega) = - \frac{\zeta_0 \tau_\Pi \omega^2}{1 + \omega^2\tau_\Pi^2}, \qquad\qquad\qquad \operatorname{Im}G^{\theta}_R(\omega) = \frac{\zeta_0 \omega}{1 + \omega^2\tau_\Pi^2}. 
\end{equation}

The imaginary part of the retarded Green's function is responsible for the \emph{dissipative} response of the system. The frequency-dependent bulk viscosity \cite{Harris:2020rus, Sawyer:1989dp, Sad:2009hba} can then be extracted,
\begin{equation}
\label{eq:kubozeta}
    \zeta(\omega) = \frac{1}{\omega} \operatorname{Im}\,G_R^\theta(\omega) = \frac{\zeta_0}{1+\omega^2 \tau_{\Pi}^2}\,,
\end{equation}
whose macroscopic counterpart is the dissipative tidal deformability $\tilde{\Xi}$ \cite{Ripley:2023lsq}.
Similarly, the real part is responsible for the reactive \emph{conservative} response, whose macroscopic counterpart is the \emph{conservative} dynamical Love number $k_l^{(2)}$ \cite{HegadeKR:2024agt}.

In the inspiral, as $T \to 0$, both $\zeta_0 \to \infty$ and $\tau_\Pi \to \infty$. Hence,
\begin{equation}
 \operatorname{Re}G^{\theta}_R(\omega) \to - \frac{\zeta_0}{\tau_\Pi}, \qquad\qquad\qquad
     \operatorname{Im}\,G_R^\theta(\omega) \to \frac{\zeta_0}{\tau_\Pi^2 \omega}. 
\end{equation}
As Eq.~\eqref{eq:transport_nsat} shows, $\frac{\zeta_0}{\tau_\Pi}$ is a constant with respect to $T$. Therefore, though the dissipative part might be significantly moderated by the extra factor of $\tau_\Pi \omega$, the conservative part will remain unchanged. Furthermore, as Fig.~1 and Fig.~3 in the main text show, $\frac{\zeta_0}{\tau_\Pi}$ can change by orders of magnitude as $L$ changes, which will lead to a significant change to $k_l^{(2)}$ \cite{HegadeKR:2024agt} and thus a very likely detectable effect. 
Equivalently, this finite real shift appears as a non-vanishing change in $k_l^{(2)}$, with the same radial weighting.
In other words, while the imaginary part governing dissipation falls as $1/\omega$ in this regime, the real part saturates at a finite value set locally by $\zeta_0/\tau_\Pi$. Propagated through the linear tidal-perturbation equations, this finite real shift produces a non-vanishing change in the dynamical Love number even if the dissipative tidal deformability is small. This clarifies why a modest dissipative signature at inspiral frequencies does not preclude an observable effect caused by bulk viscosity on the conservative tidal response.

\end{document}